\def\ra{\rangle}
\def\la{\langle}
\def\be{\begin{equation}}
\def\ee{\end{equation}}
\def\ba{\begin{array}}
\def\ea{\end{array}}

\def\Cb{{\Bbb C}}

\documentstyle[11pt]{article}
\input amssym.def
\begin{document}
\baselineskip=18pt \setcounter{page}{1} \centerline{\large\bf A Note on
Normal Forms of Quantum States and Separability} \vspace{3ex}
\begin{center}
Ming Li$^{1}$, Shao-Ming Fei$^{1,2}$ and  Zhi-Xi Wang$^{1}$

\vspace{2ex}

\begin{minipage}{5in}

\small $~^{1}$ {\small Department of Mathematics, Capital Normal
University, Beijing 100037, China}

{\small $~^{2}$ Institut f\"ur Angewandte Mathematik, Universit\"at
Bonn, D-53115, Germany}

%{\small $~^{2}$ Institut f\"ur Angewandte Mathematik, Universit\"at Bonn, D-53115}

%{\small $~^{3}$ Max-Planck-Institute for Mathematics in the Sciences, 04103 Leipzig}

\end{minipage}
\end{center}

\begin{center}
\begin{minipage}{5in}
\vspace{1.5ex} \centerline{\large Abstract} \vspace{1ex} We study
the normal form of multipartite density matrices. It is shown that
the correlation matrix (CM) separability criterion can be improved
from the normal form we obtained under filtering transformations.
Based on CM criterion the entanglement witness is further
constructed in terms of local orthogonal observables for both
bipartite and multipartite systems.

\vfill
\smallskip
\end{minipage}\end{center}
\bigskip

One of the important problems in the theory of quantum entanglement
is the separability: to decide whether or not a given quantum state
is entangled. A multipartite state $\rho_{AB\cdots C}$ is called
separable if it can be written as $\rho_{AB\cdots C}=\sum
\limits_{i}p_{i}\rho_{i}^{A}\otimes\rho_{i}^{B}\otimes\cdots\otimes\rho_{i}^{C},
$ where $\rho_{i}^{A}$, $\rho_{i}^{B},\cdots,\rho_{i}^{C}$ are
density matrices on subsystems $A$, $B,\cdots, C$, and $p_{i}\geq
0,\sum\limits_{i}p_{i}=1$. There have been many separability
criteria such as Bell inequalities \cite{bell}, PPT (positive
partial transposition) \cite{peres}, entanglement witnesses
\cite{witness,zhang}, realignment \cite{chen}, local
uncertainty relations \cite{guhne1,hofmann} etc. In
{\cite{julio}} by using the Bloch representation of density matrices
the author has presented a separability criterion, which is further
generalized to multipartite case \cite{hassan}. In \cite{leinaas}
the normal form of a bipartite state has been obtained. We indicate
that this normal form can be used to improve the separability
criteria from Bloch representation and local uncertainty relations.
In this note we study the normal form of multipartite density
matrices. We show that the correlation matrix (CM) criterion can be
improved from the normal form we obtained under filtering
transformations. Based on CM criterion we further construct the
entanglement witness in terms of local orthogonal observables (LOOs)
\cite{sixia} for both bipartite and multipartite systems.

For bipartite case, $\rho\in{\mathcal {H}}={\mathcal {H}}_{A}\otimes {\mathcal
{H}}_{B}$ with $dim\,{\mathcal {H}}_{A}=M$,
$dim\,{\mathcal {H}}_{B}=N$, $M\leq N$, is mapped to the following form
under local filtering transformations \cite{verstraete}:
\begin{eqnarray}
\label{FT}\rho\rightarrow \widetilde{\rho}=\frac{(F_{A}\otimes
F_{B})\rho(F_{A}\otimes F_{B})^{\dag}}{{\rm Tr}[(F_{A}\otimes
F_{B})\rho(F_{A}\otimes F_{B})^{\dag}]},
\end{eqnarray}
where $F_{A/B}\in GL(M/N, \Cb)$ are arbitrary invertible matrices.
This transformation is also known as stochastic local operations
assisted by classical communication (SLOCC). By the definition it is
obvious that filtering transformation will preserve the separability
of a quantum state.

It has been shown that under local filtering operations one can
transform a strictly positive $\rho$ into a normal form
\cite{leinaas},
\begin{eqnarray}
\label{NF2} \widetilde{\rho}=\frac{(F_{A}\otimes
F_{B})\rho(F_{A}\otimes F_{B})^{\dag}}{{\rm Tr}[(F_{A}\otimes
F_{B})\rho(F_{A}\otimes
F_{B})^{\dag}]}=\frac{1}{MN}(I+\sum\limits_{i=1}^{M^{2}-1}{\xi}_{i}
G_{i}^{A}\otimes G_{i}^{B}),
\end{eqnarray}
where ${\xi}_{i}\geq 0$, $G_{i}^{A}$ and $G_{i}^{B}$ are some
traceless orthogonal observables. The matrices $F_{A}$ and $F_{B}$
can be obtained by minimizing the function
\begin{eqnarray}
f(A,B)=\frac{{\rm Tr} [\rho(A\otimes B)]}{(\det A)^{1/M}(\det
B)^{1/N}},
\end{eqnarray}
where $A=F_{A}^{\dag}F_{A}$ and $B=F_{B}^{\dag}F_{B}$. In fact, one
can choose $F_{A}^{0}\equiv |\det
(\rho_{A})|^{1/2M}(\sqrt{\rho_{A}})^{-1}$, and $F_{B}^{0}\equiv
|\det (\rho_{B}^{'})|^{1/2N}(\sqrt{\rho_{B}^{'}})^{-1}$, where
$\rho_{B}^{'}={\rm Tr} _{A}(I\otimes (\sqrt{\rho_{A}})^{-1}\rho
I\otimes (\sqrt{\rho_{A}})^{-1})$. Then by the iteration one can get
the optimal A and B. In particular, there is a matlab code available
in \cite{verstraete2}. The normal form of a product state (if
exists) must be proportional to identity.

For bipartite separable states $\rho$, the CM separability criterion
\cite{julio} says that
\begin{eqnarray}
\label{CM2}||T||_{KF}\leq \sqrt{MN(M-1)(N-1)},
\end{eqnarray}
where $T$ is an $(M^{2}-1)\times (N^{2}-1)$ matrix with
$T_{ij}=MN\cdot {\rm Tr} (\rho \lambda_{i}^{A}\otimes
\lambda_{j}^{B})$, $||T||_{KF}$ stands for the trace norm of $T$,
$\lambda_{k}^{A/B}$s are the generators of $SU(M/N)$ and have been
chosen to be normalized, ${\rm Tr}
\lambda_{k}^{(A/B)}\lambda_{l}^{(A/B)}=\delta_{kl}$.

As the filtering transformation does not change the separability of
a state, one can study the separability of $\tilde{\rho}$ instead of
$\rho$. Under the normal form (\ref{NF2}) the criterion (\ref{CM2})
becomes
\begin{eqnarray}\label{g1}
\sum\limits_{i}\xi_{i}\leq\sqrt{MN(M-1)(N-1)}.
\end{eqnarray}

In \cite{guhne1} a separability criterion based on local uncertainty
relation (LUR) has been obtained. It says that for any separable
state $\rho$, \be\label{hh} 1-\sum\limits_{k}\la G_{k}^{A}\otimes
G_{k}^{B}\ra-\frac{1}{2}\la G_{k}^{A}\otimes I - I\otimes
G_{k}^{B}\ra^{2}\geq 0, \ee where $G_{k}^{A/B}$s are LOOs such as
the normalized generators of $SU(M/N)$ and $G_{k}^{A}=0$ for
$k=M^{2}+1, \cdots, N^{2}$. The criterion is shown to be strictly
stronger than the realignment criterion \cite{chen}. Under the
normal form ($\ref{NF2}$) criterion (\ref{hh}) becomes
\begin{eqnarray*}
1&-&\sum\limits_{k}\la G_{k}^{A}\otimes G_{k}^{B}\ra-\frac{1}{2}\la
G_{k}^{A}\otimes
I - I\otimes G_{k}^{B}\ra^{2} \\
&=&1-\frac{1}{\sqrt{MN}}-\frac{1}{MN}\sum\limits_{k}\xi_{k}-\frac{1}{2}(\sum\limits_{k}\la
G_{k}^{A}\ra^{2} +\sum\limits_{k}\la
G_{k}^{B}\ra^{2}-2\sum\limits_{k}\la G_{k}^{A}\ra\la
G_{k}^{B}\ra)\\
&=&1-\frac{1}{MN}\sum\limits_{k}\xi_{k}-\frac{1}{2}(\frac{1}{M}+\frac{1}{N})\geq 0,
\end{eqnarray*}
i.e.
\begin{eqnarray}\label{g2}
\sum\limits_{k}\xi_{k}\leq MN - \frac{M+N}{2}.
\end{eqnarray}
As $\sqrt{MN(M-1)(N-1)}\leq MN - \frac{M+N}{2}$ holds for any $M$
and $N$, from (\ref{g1}) and (\ref{g2}) it is obvious that the CM criterion recognizes
entanglement better when the normal form is taken into account.

We now consider multipartite systems. Let $\rho$ be a strictly
positive density matrix in ${\mathcal {H}}={\mathcal {H}}_{1}
\otimes {\mathcal {H}}_{2} \otimes \cdots \otimes {\mathcal
{H}}_{N}$, $dim\,{\mathcal {H}}_{i}=d_i$. $\rho$ can be generally
expressed in terms of the $SU(n)$ generators $\lambda_{\alpha_{k}}$
\cite{hassan},
\be\label{OS}
\ba{rcl}
\rho&=&\displaystyle\frac{1}{\Pi_{i}^{N}d_{i}}\left(\otimes_{j}^{N}I_{d_{j}}
+\sum\limits_{\{\mu_{1}\}}\sum\limits_{\alpha_{1}}
{\mathcal{T}}_{\alpha_{1}}^{\{\mu_{1}\}}\lambda_{\alpha_{1}}^{\{\mu_{1}\}}
+\sum\limits_{\{\mu_{1}\mu_{2}\}}\sum\limits_{\alpha_{1}\alpha_{2}}
{\mathcal{T}}_{\alpha_{1}\alpha_{2}}^{\{\mu_{1}\mu_{2}\}}\lambda_{\alpha_{1}}
^{\{\mu_{1}\}}\lambda_{\alpha_{2}}^{\{\mu_{2}\}}\right.\\[6mm]
&&+\sum\limits_{\{\mu_{1}\mu_{2}\mu_{3}\}}\sum\limits_{\alpha_{1}\alpha_{2}\alpha_{3}}
{\mathcal{T}}_{\alpha_{1}\alpha_{2}\alpha_{3}}^{\{\mu_{1}\mu_{2}\mu_{3}\}}\lambda_{\alpha_{1}}
^{\{\mu_{1}\}}\lambda_{\alpha_{2}}^{\{\mu_{2}\}}\lambda_{\alpha_{3}}^{\{\mu_{3}\}}\\[4mm]
&&+\cdots +\sum\limits_{\{\mu_{1}\mu_{2}\cdots\mu_{M}\}}\sum\limits_{\alpha_{1}\alpha_{2}\cdots\alpha_{M}}
{\mathcal{T}}_{\alpha_{1}\alpha_{2}\cdots\alpha_{M}}^{\{\mu_{1}\mu_{2}\cdots\mu_{M}\}}\lambda_{\alpha_{1}}
^{\{\mu_{1}\}}\lambda_{\alpha_{2}}^{\{\mu_{2}\}}\cdots\lambda_{\alpha_{M}}^{\{\mu_{M}\}}\\[3mm]
&&\left.+\cdots +\sum\limits_{\alpha_{1}\alpha_{2}\cdots\alpha_{N}}
{\mathcal{T}}_{\alpha_{1}\alpha_{2}\cdots\alpha_{M}}^{\{1,2,\cdots,N\}}\lambda_{\alpha_{1}}
^{\{1\}}\lambda_{\alpha_{2}}^{\{2\}}\cdots\lambda_{\alpha_{N}}^{\{N\}}\right),
\ea \ee where $\lambda_{\alpha_{k}}^{\{\mu_{k}\}}=I_{d_{1}}\otimes
I_{d_{2}}\otimes\cdots\otimes \lambda_{\alpha_{k}}\otimes
I_{d_{\mu_{k}+1}}\otimes\cdots\otimes I_{d_{N}}$ with
$\lambda_{\alpha_{k}}$ appears at the $\mu_k$th position and
\begin{eqnarray*}
{\mathcal{T}}_{\alpha_{1}\alpha_{2}\cdots\alpha_{M}}
^{\{\mu_{1}\mu_{2}\cdots\mu_{M}\}}=\frac{\prod_{i=1}^{M}
d_{\mu_{i}}}{2^{M}}{\rm Tr}[\rho\lambda_{\alpha_{1}}
^{\{\mu_{1}\}}\lambda_{\alpha_{2}}^{\{\mu_{2}\}}\cdots\lambda_{\alpha_{M}}^{\{\mu_{M}\}}].
\end{eqnarray*}

The generalized CM criterion says that: if $\rho$ in (\ref{OS}) is fully separable, then
\begin{eqnarray}\label{hhh}
||{\mathcal{T}}^{\{\mu_{1},\mu_{2}, \cdots, \mu_{M}\}}||_{KF}\leq
\sqrt{\frac{1}{2^{M}}\prod_{k=1}^{M}d_{\mu_{k}}(d_{\mu_{k}}-1)},
\end{eqnarray}
for $2\leq M \leq N, \{\mu_{1},\mu_{2}, \cdots, \mu_{M}\} \subset
\{1, 2, \cdots, N\}$. The KF norm is defined by
\begin{eqnarray*}
||{\mathcal{T}}^{\{\mu_{1},\mu_{2}, \cdots,
\mu_{M}\}}||_{KF}=max_{m=1, 2, \cdots,
M}||{\mathcal{T}}_{(m)}||_{KF},
\end{eqnarray*}
where ${\mathcal{T}}_{(m)}$ is a kind of matrix unfolding of
${\mathcal{T}}^{\{\mu_{1},\mu_{2}, \cdots, \mu_{M}\}}$.

The criterion (\ref{hhh}) can be improved by investigating the
normal form of (\ref{OS}).

{\sf[Theorem 1]} By filtering transformations of the form
\begin{eqnarray}
\label{LFM}
\widetilde{\rho}=F_{1}\otimes F_{2}\otimes\cdots\otimes
F_{N}\rho F_{1}^{\dag}\otimes F_{2}^{\dag}\otimes F_{N}^{\dag},
\end{eqnarray}
where $F_{i}\in GL(d_{i},\Bbb{C}), i=1, 2, \cdots N$, followed by
normalization, any strictly positive state $\rho$ can be transformed into
a normal form
\be\label{NF}
\ba{rcl}
\rho&=&\displaystyle\frac{1}{\Pi_{i}^{N}d_{i}}\left(\otimes_{j}^{N}I_{d_{j}}
+\sum\limits_{\{\mu_{1}\mu_{2}\}}\sum\limits_{\alpha_{1}\alpha_{2}}
{\mathcal{T}}_{\alpha_{1}\alpha_{2}}^{\{\mu_{1}\mu_{2}\}}\lambda_{\alpha_{1}}
^{\{\mu_{1}\}}\lambda_{\alpha_{2}}^{\{\mu_{2}\}}
+\sum\limits_{\{\mu_{1}\mu_{2}\mu_{3}\}}\sum\limits_{\alpha_{1}\alpha_{2}\alpha_{3}}
{\mathcal{T}}_{\alpha_{1}\alpha_{2}\alpha_{3}}^{\{\mu_{1}\mu_{2}\mu_{3}\}}\lambda_{\alpha_{1}}
^{\{\mu_{1}\}}\lambda_{\alpha_{2}}^{\{\mu_{2}\}}\lambda_{\alpha_{3}}^{\{\mu_{3}\}}\right.\\[6mm]
&&+\cdots
+\sum\limits_{\{\mu_{1}\mu_{2}\cdots\mu_{M}\}}\sum\limits_{\alpha_{1}\alpha_{2}\cdots\alpha_{M}}
{\mathcal{T}}_{\alpha_{1}\alpha_{2}\cdots\alpha_{M}}^{\{\mu_{1}\mu_{2}\cdots\mu_{M}\}}\lambda_{\alpha_{1}}
^{\{\mu_{1}\}}\lambda_{\alpha_{2}}^{\{\mu_{2}\}}\cdots\lambda_{\alpha_{M}}^{\{\mu_{M}\}}\\[4mm]
&&\left.+\cdots +\sum\limits_{\alpha_{1}\alpha_{2}\cdots\alpha_{N}}
{\mathcal{T}}_{\alpha_{1}\alpha_{2}\cdots\alpha_{M}}^{\{1,2,\cdots,N\}}\lambda_{\alpha_{1}}
^{\{1\}}\lambda_{\alpha_{2}}^{\{2\}}\cdots\lambda_{\alpha_{N}}^{\{N\}}\right).
\ea \ee

{\sf[Proof]} Let $D_{1}, D_{2}, \cdots, D_{N}$ be the sets of
density matrices of the $N$ subsystems. The cartesian product $D_{1}
\times D_{2} \times \cdots \times D_{N}$ consisting of all product
density matrices $\rho_{1} \otimes \rho_{2} \otimes \cdots \otimes
\rho_{N}$ with normalization ${\rm Tr} \rho_{i}=1$, $i=1, 2, \cdots,
N$, is a compact set of matrices on the full Hilbert space $\mathcal
{H}$. For the given density matrix $\rho$ we define the following
function of $\rho_{i}$
\begin{eqnarray*}
f(\rho_{1}, \rho_{2}, \cdots, \rho_{N})=\frac{{\rm Tr}
[\rho(\rho_{1}\otimes \rho_{2} \otimes\cdots\otimes
\rho_{N})]}{\prod_{i=1}^{N}\det (\rho_{i})^{1/d_{i}}}.
\end{eqnarray*}
The function is well-defined on the interior of $D_{1} \times D_{2}
\times \cdots \times D_{N}$ where $\det \rho_{i}>0$. As $\rho$ is
assumed to be strictly positive, we have ${\rm Tr}
[\rho(\rho_{1}\otimes \rho_{2} \otimes\cdots\otimes \rho_{N})]>0$.
Since $D_{1} \times D_{2} \times \cdots \times D_{N}$ is compact, we
have ${\rm Tr} [\rho(\rho_{1}\otimes \rho_{2} \otimes\cdots\otimes
\rho_{N})]\geq C>0$ with a lower bound C depending on $\rho$.

It follows that $f\rightarrow \infty$ on the boundary of $D_{1}
\times D_{2} \times \cdots \times D_{N}$ where at least one of the
$\rho_{i}$s satisfies that $\det \rho_{i}=0$. It follows further that
$f$ has a positive minimum on the interior of $D_{1} \times D_{2}
\times \cdots \times D_{N}$ with the minimum value attained for at
least one product density matrix $\tau_{1} \otimes \tau_{2} \otimes
\cdots \otimes \tau_{N}$ with $\det \tau_{i}>0$, $i=1, 2, \cdots, N$.
Any positive density matrix $\tau_{i}$ with $\det \tau_{i}>0$ can
be factorized in terms of Hermitian matrices $F_{i}$ as
\begin{eqnarray}
\label{tau}\tau_{i}=F_{i}^{\dag}F_{i}
\end{eqnarray}
where $F_{i}\in GL(d_{i}, \Bbb{C})$. Denote $F=F_{1}
\otimes F_{2} \otimes \cdots \otimes F_{N}$, so that $\tau_{1}
\otimes \tau_{2} \otimes \cdots \otimes \tau_{N} =F^{\dag}F$. Set
$\widetilde{\rho}=F\rho F^{\dag}$ and define
\begin{eqnarray*}
\widetilde{f}(\rho_{1}, \rho_{2}, \cdots \rho_{N}) &=&\frac{{\rm Tr}
[\widetilde{\rho}(\rho_{1}\otimes \rho_{2} \otimes\cdots\otimes
\rho_{N})]}{\prod_{i=1}^{N}\det (\rho_{i})^{1/d_{i}}}\\
&=&\prod_{i=1}^{N}\det (\tau_{i})^{1/d_{i}}\cdot\frac{{\rm Tr}
[\rho(F_{1}^{\dag}\rho_{1}F_{1}\otimes F_{2}^{\dag}\rho_{2}F_{2}
\otimes\cdots\otimes
F_{N}^{\dag}\rho_{N}F_{N})]}{\prod_{i=1}^{N}\det (\tau_{i})^{1/d_{i}}\det (\rho_{i})^{1/d_{i}}}\\
&=&\prod_{i=1}^{N}\det (\tau_{i})^{1/d_{i}}\cdot
f(F_{1}^{\dag}\rho_{1}F_{1}, F_{2}^{\dag}\rho_{2}F_{2}, \cdots,
F_{N}^{\dag}\rho_{N}F_{N}).
\end{eqnarray*}

We see that when $F^{\dag}_{i}\rho_{i}F_{i}=\tau_{i}$,
$\widetilde{f}$ has a minimum and
\begin{eqnarray*}
\rho_{1} \otimes \rho_{2} \otimes \cdots \otimes
\rho_{N}=(F^{\dag})^{-1} \tau_{1} \otimes \tau_{2} \otimes \cdots
\otimes \tau_{N} F^{-1}=I.
\end{eqnarray*}

Since $\widetilde{f}$ is stationary under infinitesimal variations
about the minimum it follows that
\begin{eqnarray*}
{\rm Tr}[\widetilde{\rho}\delta(\rho_{1} \otimes \rho_{2} \otimes
\cdots \otimes \rho_{N})]=0
\end{eqnarray*}
for all infinitesimal variations,
\begin{eqnarray*}
\delta(\rho_{1} \otimes \rho_{2} \otimes \cdots \otimes
\rho_{N})=\delta \rho_{1} \otimes I_{d_{2}} \otimes \cdots \otimes
I_{d_{N}}+ I_{d_{1}} \otimes \delta_{\rho_{2}} \otimes I_{d_{3}}
\otimes \cdots \otimes I_{d_{N}}\\
 + \cdots \cdots + I_{d_{1}} \otimes
I_{d_{2}} \otimes \cdots \otimes I_{d_{N-1}} \otimes \delta \rho_{N},
\end{eqnarray*}
subjected to the constraint $\det (I_{d_{i}}+\delta\rho_{i})=1$,
which is equivalent to ${\rm Tr} (\delta\rho_{i})=0$, $i=1, 2, \cdots,
N$, using $\det (e^{A})=e^{{\rm Tr} A}$ for a given
matrix $A$. Thus, $\delta\rho_{i}$ can be represented by the $SU$
generators, $\delta\rho_{i}=\sum\limits_{k}\delta
c_{k}^{i}\lambda_{k}^{i}$. It follows that ${\rm
Tr}(\widetilde{\rho}\lambda_{\alpha_{k}}^{\{\mu_{k}\}})=0$ for any
$\alpha_{k}$ and $\mu_{k}$. Hence the terms proportional to
$\lambda_{\alpha_{k}}^{\{\mu_{k}\}}$ in ($\ref{OS}$) disappear.
$\hfill\Box$

{\sf[Corollary]} The normal form of a product state in
${\mathcal {H}}$ must be proportional to the identity.

{\sf[Proof]} Let $\rho$ be such a state. From (\ref{NF}), we get
that
\begin{eqnarray}
\label{RMONF}\widetilde{\rho}_{i}={\rm Tr}_{1, 2, \cdots, i-1, i+1,
\cdots, N}\rho=\frac{1}{d_{i}}I_{d_{i}}.
\end{eqnarray}
Therefore for a product state $\rho$ we have
\begin{eqnarray*}
\rho=\rho_{1}\otimes\rho_{2}\otimes\cdots\otimes\rho_{N}=\frac{1}{\prod_{i=1}^{N}
d_{i}}\otimes_{i=1}^{N}I_{d_{i}}.
\end{eqnarray*}
$\hfill\Box$

To show the separability of multipartite states in terms of their
normal forms ($\ref{NF}$) we consider the PPT entangled edge state
\cite{acin}
\begin{eqnarray}
\rho=\left(%
    \begin{array}{cccccccc}
      1 & 0 & 0 & 0 & 0 & 0 & 0 & 1\\
      0 & a & 0 & 0 & 0 & 0 & 0 & 0\\
      0 & 0 & b & 0 & 0 & 0 & 0 & 0\\
      0 & 0 & 0 & c & 0 & 0 & 0 & 0\\
      0 & 0 & 0 & 0 & \frac{1}{c} & 0 & 0 & 0\\
      0 & 0 & 0 & 0 & 0 & \frac{1}{b} & 0 & 0\\
      0 & 0 & 0 & 0 & 0 & 0 & \frac{1}{a} & 0\\
      1 & 0 & 0 & 0 & 0 & 0 & 0 & 1\\
    \end{array}%
    \right)
\end{eqnarray}
mixed with noises:
\begin{eqnarray*}
\rho_{p}=p\rho+\frac{(1-p)}{8}I_{8}.
\end{eqnarray*}
Select $a=2, b=3$, and $c=0.6$. Using the criterion in \cite{hassan}
we get that $\rho_{p}$ is entangled for $0.92744< p \leq 1$. But
after transforming $\rho_{p}$ to its normal form (\ref{NF}), the criterion can
detect entanglement for $0.90285< p \leq 1$.

Here we indicate that the filtering transformation does not change the
PPT property. Let $\rho\in {\mathcal {H}}_{A}\otimes{\mathcal {H}}_{B}$ be PPT, i.e.
$\rho^{T_{A}}\geq 0,$ and $ \rho^{T_{B}}\geq 0$.
Let $\widetilde{\rho}$ be the normal form of $\rho$. From ($\ref{FT}$)
we have
\begin{eqnarray*}
\widetilde{\rho}^{T_{A}}=\frac{(F_{A}^{*}\otimes F_{B}) \rho^{T_{A}}
(F_{A}^{T}\otimes F_{B}^{\dag})}{{\rm Tr}[(F_{A}\otimes
F_{B})\rho(F_{A}\otimes F_{B})^{\dag}]}.
\end{eqnarray*}
For any vector $|\psi \rangle$, we have
\begin{eqnarray*}
\langle \psi | \widetilde{\rho}^{T_{A}}| \psi \rangle
&=&\frac{\langle \psi |(F_{A}^{*}\otimes F_{B}) \rho^{T_{A}}
(F_{A}^{T}\otimes F_{B}^{\dag})| \psi \rangle}{{\rm
Tr}[(F_{A}\otimes
F_{B})\rho(F_{A}\otimes F_{B})^{\dag}]}
\equiv\langle \psi^{'} |\rho^{T_{A}} |\psi^{'} \rangle\geq 0,
\end{eqnarray*}
where $|\psi^{'} \rangle=\frac{(F_{A}^{T}\otimes F_{B}^{\dag})| \psi
\rangle}{\sqrt{{\rm Tr}[(F_{A}\otimes F_{B})\rho(F_{A}\otimes
F_{B})^{\dag}]}}.$ $\widetilde{{\rho}}^{T_{B}}\geq 0$ can be proved
similarly. This property is also valid for multipartite case.
Hence a bound entangled state will be bound entangled under
filtering transformations.

For N-partite systems in ${\mathcal {H}}={\mathcal
{H}}_{1}\otimes {\mathcal {H}}_{2}\otimes\cdots\otimes {\mathcal
{H}}_{N}$ $(N\geq 2)$ with $dim\,{\mathcal {H}}_{i}=d_{i}$, $i=1, 2,
\cdots, N$, the local orthogonal observables (LOOs) can be given in
the following way. Assume $d_n\equiv \max\{d_{i}, i=1, 2, \cdots,
N\}$. One can choose $d^{2}$ observables
$G_{k}^{n}$ associated with the subsystem ${\mathcal {H}}_{n}$. For other subsystems
with smaller dimensions, say ${\mathcal {H}}_{1}$, one can choose
$d_{1}^{2}$ observables $G_{k}^{1}$, $k=1,2,\cdots, d_{1}^{2}$ and set
$G_{k}^{1}=0$ for $k=d_{1}^{2}+1, \cdots, d^{2}$. Based on CM
criterion we can further construct entanglement witness (EW) in terms of
such LOOs. EW \cite{sixia} is an
observable of the composite system that has (i) nonnegative
expectation values in all separable states and (ii) at least one
negative eigenvalue (or equivalently, can recognizes at least one
entangled state).

We first consider bipartite systems in ${\mathcal {H}}_{A}^{M}\otimes
{\mathcal {H}}_{B}^{N}$ with $M\leq N$.

{\sf[Theorem 2]} For any LOOs $G^{A}_{k}$ and $G^{B}_{k}$,
$$
W=I-\alpha\sum\limits^{N^{2}-1}_{k=0}G^{A}_{k}\otimes
G^{B}_{k}
$$
is an EW, where
$\alpha=\frac{\sqrt{MN}}{\sqrt{(M-1)(N-1)}+1}$ and
\begin{eqnarray}\label{th2c}
G^{A}_{0}=\frac{1}{\sqrt{M}}I_{M},~~~
G^{B}_{0}=\frac{1}{\sqrt{N}}I_{N}.
\end{eqnarray}

{\sf[Proof]} Let $\rho=\sum\limits_{l,m=0}^{N^{2}-1}T_{lm}\lambda_{l}^{A}\otimes\lambda_{m}^{B}$
be a separable state,
where $\lambda_{k}^{A/B}$ are normalized generators of $SU(M/N)$ with
$\lambda^{A}_{0}=\frac{1}{\sqrt{M}}I_{M}$,
$\lambda^{B}_{0}=\frac{1}{\sqrt{N}}I_{N}$. Any other LOOs
$G^{A/B}_{k}$ fulfill ($\ref{th2c}$) can be obtained from these
$\lambda$s through orthogonal transformations ${\mathcal{O}}^{A/B}$,
$G^{A/B}_{k}=\sum\limits_{l=0}^{N^{2}-1}{\mathcal
{O}}^{A/B}_{kl}\lambda_{l}$, where
${\mathcal{O}}^{A/B}=\left(%
    \begin{array}{cc}
      1 & 0  \\
      0 & {\mathcal {R}}^{A/B} \\
    \end{array}%
    \right)$,
${\mathcal {R}}^{A/B}$ are $(N^{2}-1)\times (N^{2}-1)$
orthogonal matrices. We have
\begin{eqnarray*}
{\rm Tr}  \rho W&=&1-\alpha \frac{1}{\sqrt{MN}}- \alpha
\sum_{k=1}^{N^{2}-1}\sum_{l,m=1}^{N^{2}-1}{\mathcal {R}}_{kl}^{A}{\mathcal {R}}_{km}^{B}
{\rm Tr}  \rho(\lambda^{A}_{l}\otimes \lambda^{B}_{m})\\
&=&\frac{\sqrt{(M-1)(N-1)}}{\sqrt{(M-1)(N-1)}+1}-\frac{1}{\sqrt{MN}(\sqrt{(M-1)(N-1)}+1)}
\sum_{k=1}^{N^{2}-1}\sum_{l,m=1}^{N^{2}-1}{\mathcal {R}}_{kl}^{A}T_{lm}{\mathcal {R}}_{km}^{B}\\
&=&\frac{\sqrt{(M-1)(N-1)}}{\sqrt{(M-1)(N-1)}+1}-\frac{1}{\sqrt{MN}(\sqrt{(M-1)(N-1)}+1)}
{\rm Tr} ({\mathcal {R}}^{A}T({\mathcal {R}}^{B})^{T})\\[2mm]
&\geq&\frac{\sqrt{MN(M-1)(N-1)}-||T||_{KF}}{\sqrt{MN}(\sqrt{(M-1)(N-1)}+1)}\geq
0,
\end{eqnarray*}
where we have used ${\rm Tr}  ({\mathcal
{R}}T) \leq ||T||_{KF}$ for any unitary ${\mathcal {R}}$
in the first inequality and the CM criterion in the second inequality.

Now let
$\rho=\frac{1}{MN}(I_{MN}+\sum\limits_{i=1}^{M^{2}-1}s_{i}\lambda_{i}^{A}\otimes
I_{N}+\sum\limits_{j=1}^{N^{2}-1}r_{j}I_{M}\otimes\lambda_{j}^{B}+\sum\limits_{i=1}^{M^{2}-1}
\sum\limits_{j=1}^{N^{2}-1}T_{ij}\lambda_{i}^{A}\otimes
\lambda_{j}^{B})$ be a state in ${\mathcal
{H}}_{A}^{M}\otimes{\mathcal {H}}_{B}^{N}$  which
violates the CM criterion. Denote $\sigma_{k}(T)$ the singular values
of $T$. By singular value decomposition, one has $T= U^{\dag}
\Lambda V^{*}$, where $\Lambda$ is a diagonal matrix with
$\Lambda_{kk}=\sigma_{k}(T)$. Now choose LOOs to be
$G_{k}^{A}=\sum_{l}U_{kl}\lambda^{A}_{l}$,
$G_{k}^{B}=\sum_{m}V_{km}\lambda^{B}_{m}$ for $k=1,2,\cdots,N^{2}-1$
and $G_{0}^{A}=\frac{1}{M}I_{M}, G_{0}^{B}=\frac{1}{N}I_{N}$. We
obtain
\begin{eqnarray*}
{\rm Tr}  \rho W&=&1-\alpha \frac{1}{\sqrt{MN}}- \alpha
\sum_{k=1}^{N^{2}-1}\sum_{l,m=1}^{N^{2}-1}U_{kl}V_{km}
{\rm Tr}  \rho(\lambda^{A}_{l}\otimes \lambda^{B}_{m})\\
&=&\frac{\sqrt{(M-1)(N-1)}}{\sqrt{(M-1)(N-1)}+1}-\frac{1}{\sqrt{MN}(\sqrt{(M-1)(N-1)}+1)}
{\rm Tr} (UTV^{T})\\
&=&\frac{\sqrt{MN(M-1)(N-1)}-||T||_{KF}}{\sqrt{MN}(\sqrt{(M-1)(N-1)}+1)}<
0
\end{eqnarray*}
where the CM criterion has been used in the last step. \hfill$\Box$

As the CM criterion can be generalized to multipartite form in
\cite{hassan}, we can also define entanglement witness for
multipartite system in ${\mathcal {H}}_{1}^{d_{1}}\otimes {\mathcal
{H}}_{2}^{d_{2}}\otimes\cdots\otimes {\mathcal {H}}_{N}^{d_{N}}$.
Set $d(M)=\max\{d_{\mu_{i}}, i=1, 2, \cdots, M\}$. Choose LOOs
$G^{\{\mu_{i}\}}_{k}$ for $0\leq k\leq d(M)^{2}-1$ with
$G^{\{\mu_{i}\}}_{0}=\frac{1}{d_{\mu_{i}}}I_{d_{\mu_{i}}}$ and
define
\begin{eqnarray}
\label{EWFM}
W^{(M)} = I-\beta^{(M)}
\sum_{k=0}^{d(M)^{2}-1}G_{k}^{\{\mu_{1}\}}\otimes
G_{k}^{\{\mu_{2}\}}\otimes \cdots \otimes G_{k}^{\{\mu_{M}\}},
\end{eqnarray}
where
$\beta^{(M)}=\frac{\sqrt{\prod_{i=1}^{M}d_{\mu_{i}}}}{1+\sqrt{\prod_{i=1}^{M}(d_{\mu_{i}}-1)}},
2 \leq M\leq N$. One can prove that ($\ref{EWFM}$) is an EW
candidate for multipartite states. First we assume
$||{\mathcal{T}}^{(M)}||_{KF}=||{\mathcal{T}}_{(m_{0})}||_{KF}$.
Note that for any ${\mathcal{T}}_{(m_{0})}$, there must exist an
elementary transformation $P$ such that
$\sum\limits_{k=1}^{d(M)^{2}-1}{\mathcal{T}}_{kk\cdots
k}^{\{\mu_{1}\mu_{2}\cdots\mu_{M}\}} ={\rm Tr}
({\mathcal{T}}_{(m_{0})}P)$. Then for an N-partite separable state
we have
\begin{eqnarray*}
{\rm Tr}  \rho W^{(M)}&=&1-\beta^{(M)}
\frac{1}{\sqrt{\prod_{i=1}^{M}d_{\mu_{i}}}}-\beta^{(M)}
\frac{1}{\prod_{i=1}^{M}d_{\mu_{i}}}\sum\limits_{k=1}^{d(M)^{2}-1}{\mathcal{T}}_{kk\cdots
k}^{\{\mu_{1}\mu_{2}\cdots\mu_{M}\}}\\
&=&1-\beta^{(M)}
\frac{1}{\sqrt{\prod_{i=1}^{M}d_{\mu_{i}}}}-\beta^{(M)}
\frac{1}{\prod_{i=1}^{M}d_{\mu_{i}}}{\rm Tr} ({\mathcal{T}}_{(m_{0})}P)\\
&\geq&1-\beta^{(M)}
\frac{1}{\sqrt{\prod_{i=1}^{M}d_{\mu_{i}}}}-\beta^{(M)}
\frac{1}{\prod_{i=1}^{M}d_{\mu_{i}}}||{\mathcal{T}}_{(m_{0})}||_{KF}\\
&\geq&1-\beta^{(M)}
\frac{1}{\sqrt{\prod_{i=1}^{M}d_{\mu_{i}}}}-\beta^{(M)}
\frac{1}{\prod_{i=1}^{M}d_{\mu_{i}}}\sqrt{\prod_{k=1}^{M}d_{\mu_{k}}(d_{\mu_{k}}-1)}\\
&=&0
\end{eqnarray*}
for any $2\leq M\leq N$, where we have used that $P$ must be
orthognal matrix and ${\rm Tr} (MU)\leq ||M||_{KF}$ for any unitary
$U$ at the first inequality. The second inequality is due to the
generalized CM criterion.

By choosing proper LOOs it is also easy to show that $W^{(M)}$
has negative eigenvalues. For example for three qubits case , 
taking the normalized pauli matrices as LOOs, one find a negative eigenvalue
of $W^{(M)}$, $\frac{1-\sqrt{3}}{2}$.

We have studied the normal form of multipartite density matrices.
It has been shown that separability criteria
can be improved by transforming the states to their normal forms through filtering
transformations. The entanglement witness
has been constructed in terms of local orthogonal observables
for both bipartite and multipartite systems.
Here we considered only the strictly positive (full rank) states.
Although full rank is a sufficient condition for the existence of the normal forms,
in fact for many rank deficiency density matrices their normal forms can be also calculated.

\bigskip
\noindent{\bf Acknowledgments}\, This work is supported by the NSFC
10675086, NSF of Beijing 1042004 and KM200510028022, NKBRPC(2004CB318000).


\smallskip
\begin{thebibliography}{99}
\bibitem{bell} J. S. Bell, Physics (Long Island City, N.Y.) {\bf 1}, 195 (1964).

\bibitem{peres} A. Peres, Phys. Rev. Lett. 77, 1413 (1996); M. Horodecki,
P. Horodecki, and R. Horodecki, Phys. Lett. A 223, 1 (1996).

\bibitem{witness} M. Lewenstein, B. Kraus, J. I. Cirac, and P.
Horodecki, Phys. Rev. A 62, 052310 (2000); D. Bruss, J. I. Cirac, P.
Horodecki, F. Hulpke, B. Kraus, M. Lewenstein, and A. Sanpera, J.
Mod. Opt 49, 1399 (2002).

\bibitem{zhang} C. J. Zhang, Y. S. Zhang, S. Zhang, and G. C. Guo,
Phys. Rev. A 76, 012334 (2007).

\bibitem{chen} K. Chen and L. A. Wu. Quantum Inf. Comput. {\bf 3}, 193(2003).

\bibitem{guhne1} O. G\"{u}hne, M. Mechler, G. T¨®th  and P.
Adam, Phys. Rev. A {\bf 74}, 010301(R)(2006).

\bibitem{hofmann} H. F. Hofmann and S. Takeuchi, Phys, Rev. A {\bf 68}, 032103 (2003).

\bibitem{julio} J. I. de Vicente, Quantum Inf. Comput. {\bf 7}, 624 (2007).

\bibitem{hassan} A. Saif M. Hassan and Pramod S. Joag, quant-ph/0704.3942 (2007).

\bibitem{leinaas} J. M. Leinaas, J. M. and E. Ovrum, Phys, Rev. A {\bf 74}, 012313 (2006).

\bibitem{sixia} S. X. Yu and N. L. Liu, Phys. Rev. Lett. {\bf 95}, 150504 (2005).

\bibitem{verstraete} F. Verstraete, J. Dehaene,
and B. De Moor, Phys. Rev. A {\bf 68}, 012103 (2003).

\bibitem{verstraete2} F. Verstraete, Ph. K. Thesis, Katholieke
Universiteit Leuven (2002).

\bibitem{acin} A. Ac\'{i}n, K. Bru{\ss}, M. Lewenstein and A.
Sanpera, Phys. Rev. Lett. {\bf 87}, 040401 (2001).

\end{thebibliography}
\end{document}